\documentclass[%
 reprint,
superscriptaddress,
 amsmath,amssymb,
 aps,
 pre,
floatfix,
]{revtex4-2}

\usepackage{pdfpages} 
\usepackage{pgffor} 

\makeatletter
\AtBeginDocument{\let\LS@rot\@undefined}
\makeatother

\def\supplementfilename{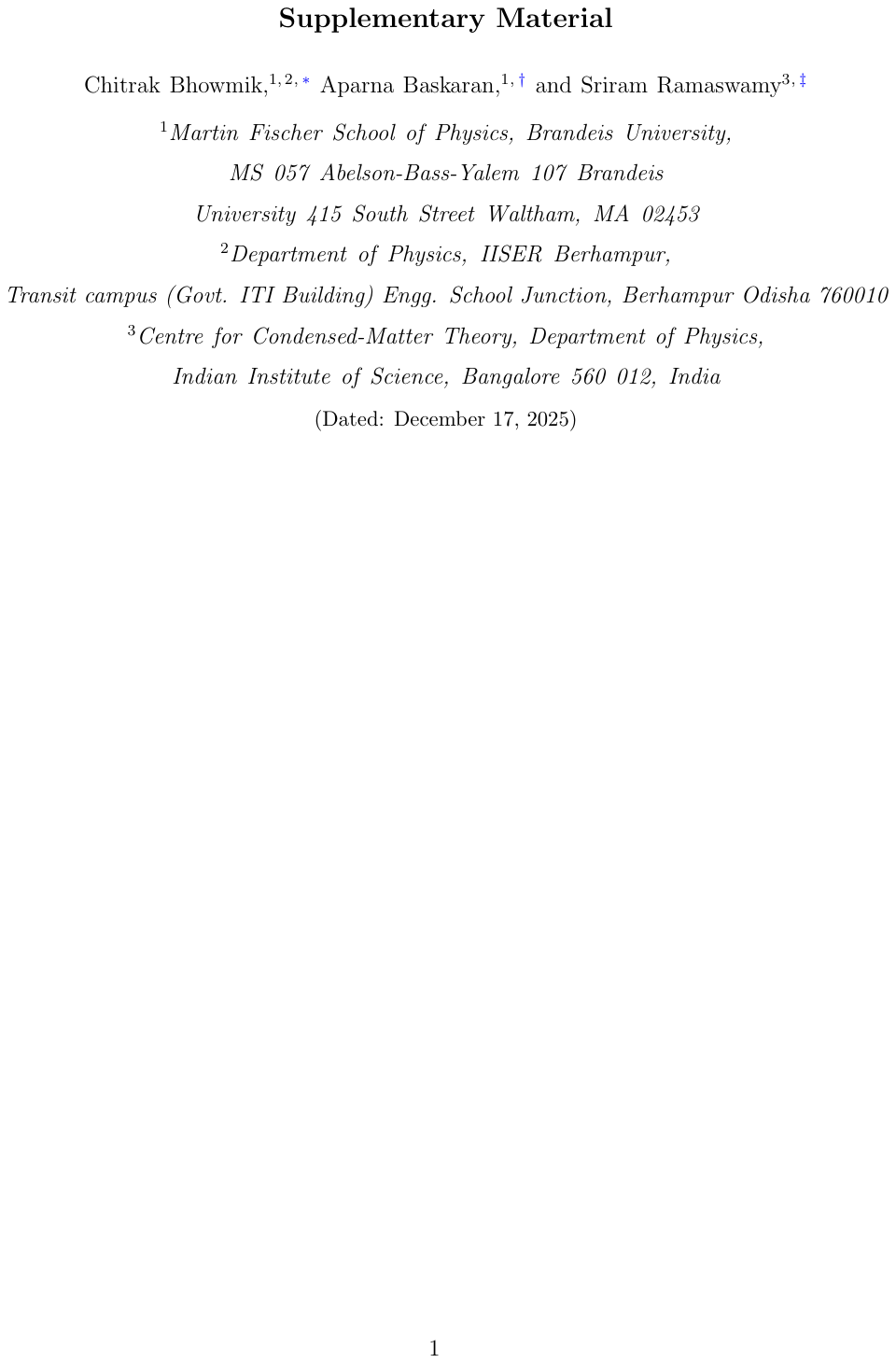}

\pdfximage{\supplementfilename}
\def\numbersupplementpages{\the\pdflastximagepages}

\newif\ifarXiv
\arXivtrue 

\usepackage{hyperref}
\hypersetup{
    colorlinks,
    linkcolor={blue},
    citecolor={blue},
    urlcolor={blue!80!black}
}
\usepackage{ulem}
\usepackage{xr}
\usepackage{xcolor}
\usepackage{braket}
\usepackage{graphicx}
\usepackage{dcolumn}
\usepackage{bm}

\begin{document}

\title{Segregation dynamics in active-passive
mixtures of semiflexible filaments}

\author{Chitrak Bhowmik}
\email{chitrak@brandeis.edu}
\affiliation{Martin Fischer School of Physics, Brandeis University, MS 057
Abelson-Bass-Yalem 107
Brandeis University
415 South Street
Waltham, MA 02453}
\affiliation{Department of Physics, IISER Berhampur, Transit campus (Govt. ITI Building)
Engg. School Junction, Berhampur
Odisha 760010}
\author{Aparna Baskaran}%
 \email{aparna@brandeis.edu}
\affiliation{Martin Fischer School of Physics, Brandeis University, MS 057
Abelson-Bass-Yalem 107
Brandeis University
415 South Street
Waltham, MA 02453}

\author{Sriram Ramaswamy}
 \homepage{sriram@iisc.ac.in}
\affiliation{Centre for Condensed Matter Theory, Department of Physics, Indian Institute of Science, Bangalore 560 012, India}%

\date{\today}

\begin{abstract}
We study the segregation of motile semiflexible filaments from a background of similar but non-motile filaments. Our Langevin dynamics simulations reveal a wide range of emergent structures governed by filament flexibility and activity, i.e., self-propulsion strength. The system segregates at low activities, while at high activities it undergoes remixing which is a characteristic feature of semi-flexible active filaments. We show that collision-induced softening of single filaments is the dominant mode for this remixing. We provide a scaling argument for the lowering of the active polymer stiffness and show that it agrees well with the lowering of the segregation order parameter. We expect that our studies will shed light on the spatial organization of  biofilaments within the cell, on the plasma-membrane, and beyond, and help in the design of novel biomaterials whose structure can be tuned via the properties of the active or the passive filaments.
\end{abstract}

\maketitle


\section{\label{sec:Introduction}Introduction}

Spatially heterogeneous structure in fluid mixtures at thermal equilibrium generally requires the presence of a mediating agent such as an amphiphile, or a long-range constraint such as the connectivity of block copolymers which prevents macroscopic demixing \cite{Leibler1980,Carrre2023}. Far from equilibrium, reaction–diffusion systems can also destabilize uniform states and produce stationary or oscillatory patterns via Turing-like instabilities \cite{Landge2020}. Heterogeneity in fluids is ubiquitous in biology. Important examples include mesoscale structure in cell membranes \cite{saha2022active,Simons2011,NietoGarai2022}, biomolecular condensates in cells \cite{Shin2017,Brangwynne2015,Feric2016}, and segregation in nuclei \cite{Mahajan2022, Zidovska2013} which are crucial for the functioning of cells. The inherent non-equilibrium nature of biological systems makes these possible.

\par 

The tools from the field of active matter have proved to be very successful in explaining such nonequilibrium organization \cite{Stenhammar2015,Negi2025,Agrawal2021,Grosberg2015,Weber2016,Chattopadhyay2023}. Chromosomal segregation within the nucleus has been explained by considering the activity differences between sections of the genome\cite{Ganai2014}. Differential motility has been found to be responsible for segregation in keratocyte colonies \cite{Mhes2012,Belmonte2008}. Enzymatic differences between mixtures have been shown to lead to multi-phase separation \cite{Monahan2017,Chen2024}. Binary mixtures of molecular motors strongly segregate in microtubule motility assays \cite{utzschneider2024force}, through a mechanism closely related to that operating in artificial motile/non-motile mixtures \cite{kant2024bulk}. The common thread in these models is the presence of two kinds of species with differential activity.

\par 

Mixtures of active and passive objects have been shown to phase-separate.  \cite{Weber2016,Dolai2018,McCandlish2012,Smrek2017,Stenhammar2015}. Existing works have looked at multiple ways of pushing these mixtures out of equilibrium, which can be broadly classified into two : two-temperature models (where both species execute thermal Brownian motion but one is held at a higher temperature than the other and defined as active) \cite{Grosberg2015,Weber2016,Smrek2017,chattopadhyay2021heating} and vectorial models (where the active particles are driven with a constant speed) \cite{Stenhammar2015,Dolai2018,McCandlish2012,Kumar2018}. Although a lot has been uncovered about the physical principles that drive segregation of such mixtures, the effect of the internal degrees of freedom on the collective dynamics has been largely unexplored. Self-propelled semi-flexible polymers serve as a good model to study the effects of active forces and torques on the filament conformation which include coiling and buckling. \cite{IseleHolder2015,Chelakkot2014,Bourdieu1995}. Collections of semiflexible-polymers have also shown interesting collective behaviour including the loss in flocking as a function of activity \cite{Kumar2018,Duman2018,Prathyusha2018,Joshi2019}. Mixtures of self-propelled semiflexible polymers can therefore serve as an idealized model to explore the role of individual units in driving segregation.

\par 

In this work, we use Langevin dynamics simulations to study the segregation dynamics of a 2D mixture of active and passive semi-flexible filaments. We map out the segregation dynamics as a function of activity and polymer stiffness. We find a non-monotonic dependence of segregation on activity which can be explained by a collision-induced softening of the polymers. A simple scaling relation obtained by equating the filament bending modulus to active stresses provides a good fit to the data. 
We provide design principles that can help choose activity and filament stiffness, which can be harnessed to engineer bespoke composite active materials \cite{Needleman2017}. Our work finds direct relevance in motility assay experiments \cite{Sciortino2021,Sciortino2022,Sumino2012,Farhadi2018,Kuera2022} and actin-microtubule composites \cite{Farhadi2018,Sciortino2022}. 

\par 

The layout of the paper is as follows. In Sec. \ref{sec:Model and Methods} we introduce the coarse-grained model of semi-flexible filaments we are implementing to simulate the active-passive mixtures. We introduce the segregation order parameter which will be used throughout the paper as a measure of the extent of mixing. Sec. \ref{sec:Results} consists of detailed descriptions of our findings, namely, how active and passive filament properties affect the segregation dynamics. We end in Sec. \ref{sec:Discussions} by discussing the results obtained in this work and listing some possible experiments for which our simulations will be useful.

\section{\label{sec:Model and Methods}Model and Methods}

We study mixtures of active and passive filaments using Langevin dynamics simulations. Each of the filaments is modeled by the active Wormlike-Chain model \cite{doi1988theory,IseleHolder2015} the dynamics for which are given by a set of $N$ Langevin equations, one for each bead. The Langevin equation for the $i$th bead of a filament ($i$ from 1 to $N$) is given by 

\begin{align}
        m_i \frac{\partial^2 \textbf{r}_i}{\partial t^2} = -\gamma\frac{\partial \textbf{r}_i}{\partial t} - \bm{\nabla} U_\text{tot} + \textbf{f}_{a} + \boldsymbol{\eta}
\end{align}
where $\textbf{r}_i$ is the position vector of the $i$th bead of a filament, $m_i$ is the mass of each bead which is set to a value of 1 and $\gamma$ is the damping coefficient.

The forces on each bead are given by the gradient of the total potential $U_\text{tot} = U_{\text{stretch}}+U_{\text{bend}} +U_{\text{excluded-volume}}$ which includes appropriate stretching, bending and non-bonding interaction terms. 
\\[6 pt]
1. $U_{\text{stretch}} = \frac{1}{2}k_s (r_{i,i+1}-R_0)^2
$ is a harmonic bond potential that encodes chain stretching where $k_s$ is the spring constant and $R_0$ is the equilibrium bond length. We define $r_{ij} = |\textbf{r}_i - \textbf{r}_j|$ to be the distance between two beads at positions $\textbf{r}_i$ and $\textbf{r}_j$. We set $k_s$ to be 5000 $k_BT$ which effectively makes the filaments inextensible. 
\\[6 pt]
2. We introduce semi-flexibility using a harmonic bending potential $ U_{\text{bend}} = \kappa^{\text{a,p}} (\theta - \pi)^2$,where $\kappa^{\text{a,p}}$ is the bending stiffness of the active and the passive filaments respectively, and $\theta$ is the angle between neighboring bonds.
\\[6 pt]
3. Beads that are not bonded interact with each other by a soft repulsion governed by the Weeks-Chandler-Andersen Potential
    \begin{align}
        U_{\text{WCA}}(r_{ij}) = 
     \begin{cases}
         4\epsilon\left[\left(\frac{\sigma}{r_{ij}}\right)^{12} - \left(\frac{\sigma}{r_{ij}}\right)^{6} + \frac{1}{4}\right] & , \, r_{ij}<2^{1/6}\sigma \\
         0 &, \, r_{ij}>2^{1/6}\sigma 
     \end{cases}
     \label{wca}
    \end{align}
where $\epsilon$ and $\sigma$ set the energy and length scales. The potential ensures an effective diameter of $\sigma$ for each bead. Time is measured in units of $\tau = \sqrt{m\sigma^2 /\epsilon}$. We set the timestep to be $10
^{-3}\tau$ in the numerical simulations.

The active forces ($\textbf{f}_a$) are introduced as forces along bond unit-vectors ($(\textbf{r}_{i}-\textbf{r}_{i+1})/r_{i,i+1}$) and are equally distributed among the beads constituting a bond. This form of activity simulates tangential driving in motility assay experiments. Thermal noise is introduced as a Gaussian random white noise
$\boldsymbol{\eta}$ satisfying the fluctuation-dissipation relation $\braket{\eta^i_\alpha (0)\eta^j_\beta (t)} = 2\gamma k_B T \delta_{\alpha \beta}\delta_{ij} \delta(t)$.
\begin{figure}[h]
    \centering
    \includegraphics[width=8.6cm]{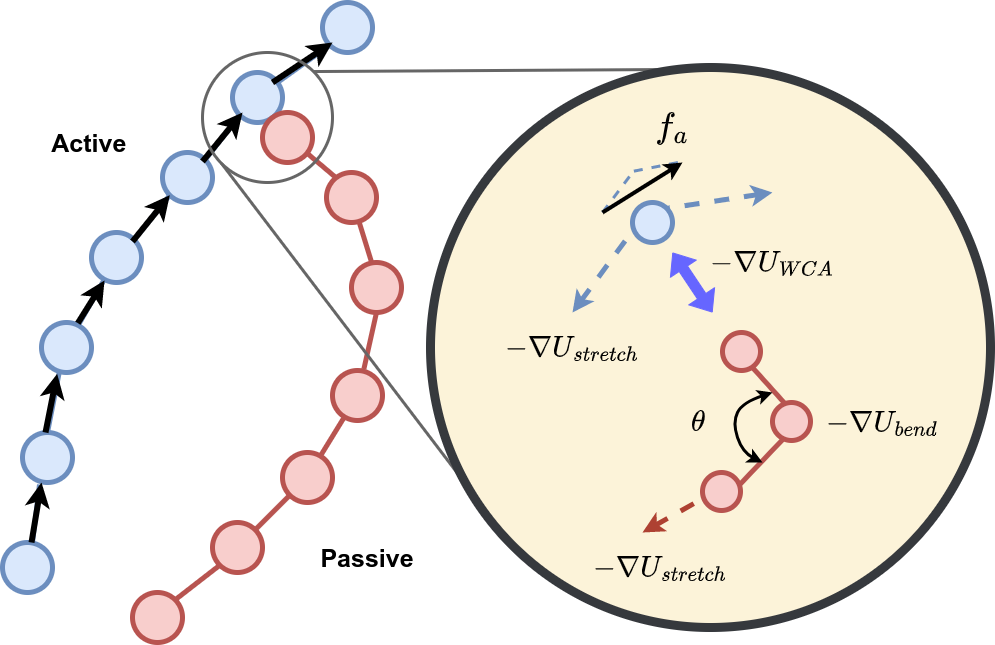}
    \caption{Schematic representation of our polymer model (both active and passive) showing all relevant forces involved. Harmonic bonds join together each bead constituting a filament. The conservative forces are shown as gradients of appropriate potentials while the non-conservative active force acts along bond unit vectors which leads to an effective tangent force along a bead.}
    \label{fig:model}
\end{figure}

\begin{figure*}[t]
    \centering
    \includegraphics[width=\linewidth]{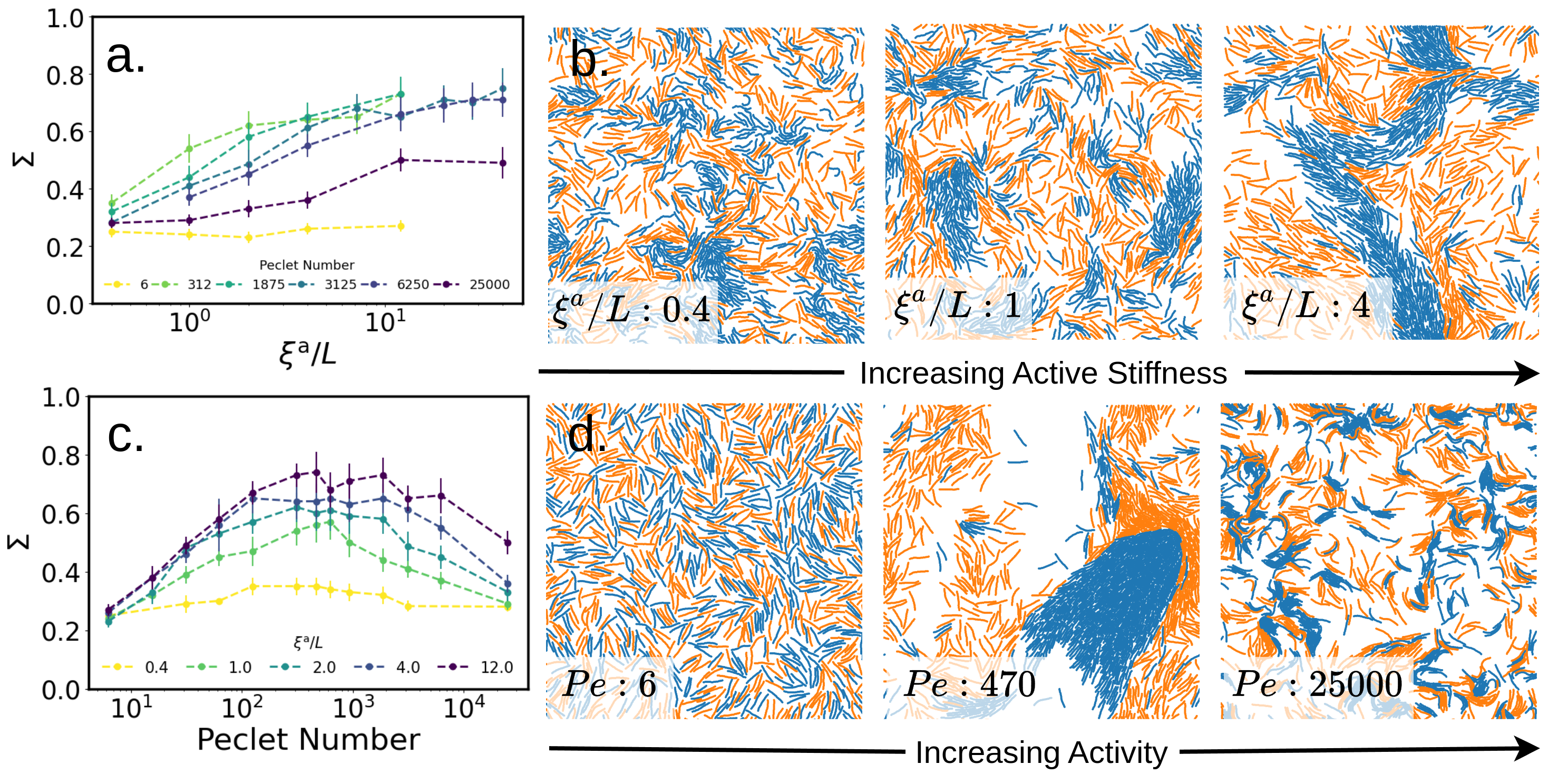}
    \caption{ \textbf{Dependence of segregation on activity and active filament stiffness.}(a) Plot showing the variation of the segregation order parameter as a function of the active filament stiffness. 
    (b) Simulation snapshots of the polymer mixtures at steady state for increasing active filament stiffness. Stiffer active filaments lead to better segregation. For these snapshots Pe is set to 125. (c) Plot showing the variation of the segregation order parameter as a function of activity. Active stiffness $\xi^a$ is set to 4. (d) Simulation snapshots of the polymer mixtures at steady state for increasing activity. For all runs passive stiffness was kept fixed at $\xi^\text{p} : 4$}
    \label{fig:fig1}
\end{figure*}

The equations of motion are integrated using a velocity-verlet integrator in LAMMPS \cite{Thompson2022}. For a typical simulation, a collection of 1000 (500 active and 500 passive) filaments of length $L=25\sigma$ was initialized in a nematically aligned state in the simulation box ($L_{\text{box}} = 200\sigma$), after which it was relaxed by running a short trajectory with a temperature ramp from 1 to 10 $k_BT$ for $2500\tau$ and thermalised by running a trajectory with a temperature ramp from 10 to 1 $k_B T$ for $10^6 \tau$ followed by a production run at 1 $k_B T$ with the activity turned on until the system reached steady state. Production run times vary from $10^7 \tau$ to $10^8 \tau$ depending on the activity. \cite{Supplement1}
\\[6 pt]
\emph{How mixed is well mixed? - Segregation analysis - }Our main goal of this study is to understand the segregation dynamics of the mixture of active-passive semi-flexible filaments. For that purpose, we need to define an order parameter that characterizes how well-mixed the two components are. We divide our simulation box into N square cells of side-length \textit{a} and calculate an order parameter ($\Sigma$) as defined by
\begin{align}
    \Sigma (a) = \frac{1}{2\phi^{(a)}(1 - \phi^{(a)})}\sum_{i=1}^N \left(\frac{n_i}{n_\text{tot}}\right)|\phi_i - \phi^{(a)}|
\end{align}
where $\phi_i$ is the fraction of active filaments, $n_i$ is the total number of filaments in the $i$th cell, $n_\text{tot}$ is the total number of filaments in the simulation box and $\phi^{(a)}$ is the total fraction of active filaments in the simulation box \cite{McCandlish2012}.

\par 

 If the mixture is homogeneous, the average active fraction is the same over all the small cells. The parameter essentially gives the deviation of the active fraction from the well-mixed limit for each cell weighted by the fraction of total filaments in a cell. The segregation parameter ($\Sigma$) is dependent on the cell size ($a$), making it necessary for us to choose a proper value of $a$. $\Sigma$ decreases monotonically with increasing the cell size. We divide the simulation box into a $9\times 9$ set of cells, which ensures that the cell side length is larger than the bead diameter but much smaller than the simulation box size.
 
\par 

\textit{Tunable Parameters} - We map the segregation order parameter ($\Sigma$) as a function of two dimensionless quantities, the Peclet number $Pe = \frac{|\textbf{f}_a| L^2}{\sigma k_B T}$, a measure of activity, and the scaled persistence lengths $\xi^{\text{a},\text{p}} /L$, which characterize filament stiffness.

\section{\label{sec:Results}Results}

\subsection{Segregation Dynamics : Strange re-mixing at High Peclet}

We study the segregation dynamics as a function of 
$\xi^{\text{a},\text{p}}$ and $Pe$. 
We find that increasing $\xi^{\text{a},\text{p}}$ leads to a monotonic increase in $\Sigma$ (Fig. \ref{fig:fig1}a). In contrast, increasing $Pe$ shows an initial increase in segregation and then a re-entrance at high $Pe$ where the mixture becomes homogeneous (Fig. \ref{fig:fig1}b). The disparity between the frequency of active-active and active-passive rod collisions renders the homogeneous mixture unstable and leads to segregation \cite{McCandlish2012}. At low activities, the mixture remains homogeneous as the self-propulsion is not strong enough to lead to alignment and clustering. At intermediate activities ($10<Pe<1000$) one observes a highly segregated state with large active flocks (Fig. \ref{fig:fig1}d). The mechanism for the formation of the flocks has been well studied \cite{Peruani2006,Toner1998,Schaller2010,Peruani2010,Baskaran2008} and can be understood as the result of collisions between active rods which leads to alignment and clustering. 
Interestingly, we find that in this highly segregated state the passive filaments also show aligned structures which are organized by the active flocks. This has been observed in motility assay experiments of actin-microtubule composites where propelled microtubules aid in the organization of the actin network \cite{Kuera2022}. Upon further increase in activity we see that the segregation starts decreasing.  This re-entrance is unique to semi-flexible filaments \cite{Prathyusha2018,Duman2018} and is not observed in active Brownian particles or stiff hard rods.

\subsection{Effective stiffness drives segregation: Polymer persistence analysis}

\begin{figure}[h]
    \centering
    \includegraphics[width=\linewidth]{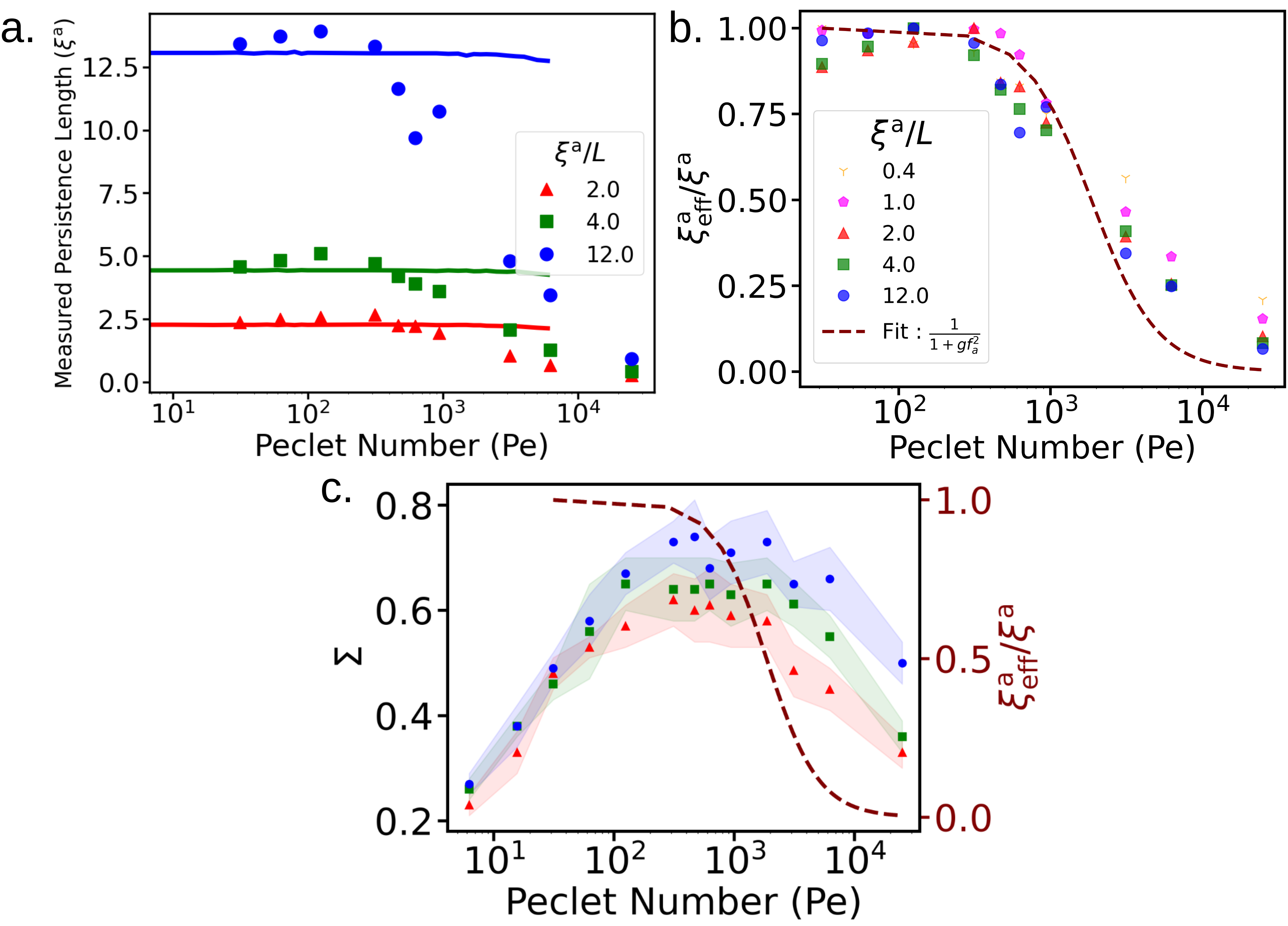}
    \caption{\textbf{Collision driven filament bending explains the loss in segregation} (a) Plot comparing the decrease in persistence lengths of isolated active filaments (solid lines) and active filaments in mixtures (individual dots). The persistence length decrease for active filaments in mixtures is significantly more as compared to isolated filaments.(b) Plot showing the scaled persistence length as a function of activity. The curves for different active stiffnesses collapse onto a single curve and can be explained by a simple scaling argument based on collisions. The single fit parameter (g) is 0.11. (c) Plot showing the change in the segregation order parameter as a function of activity for different active stiffness values ($\xi^a /L \in \{2,4,12\}$). The effective stiffness (in maroon) is overlayed on top of the plot.}
    \label{fig:fig2}
\end{figure}

To understand the cause behind the remixing at higher Peclet, we looked at how a single filament behaves under increasing Peclet. We find that increasing activity leads to a significant loss in filament stiffness (Fig.~\ref{fig:fig2}a). We quantify stiffness by calculating the persistence length of a filament. The persistence length of a filament is obtained by calculating the tangent-tangent correlations $\braket{\hat{t}(0).\hat{t}(s)}$ and fitting it to an exponential of the form $e^{-s/\xi^{\text{a}}_m}$, where s is the distance along the contour and the fit parameter $\xi^{\text{a}}_m$ is the measured persistence length. The effect we see here is different from the coiling-induced crumpling observed in simulations of individual active polymers \cite{IseleHolder2015}. Fig.~\ref{fig:fig2}a shows that the persistence lengths of active polymers in mixtures decrease significantly as a function of activity compared to a single polymer, suggesting that collisions play a role in this. We understand this softening by a simple collision frequency based scaling argument \cite{Joshi2019}. The effective stiffness is determined by the competition between elastic stresses and collision-induced stresses. The elastic stresses scale linearly with the filament bending stiffness ($\sim \kappa$). The collisional stresses occur due to both passive and active collisions. The passive stresses are order 1 while the active stresses will depend on the rate of momentum transfer which scales as the square of the active force ($\sim |\textbf{f}_a| ^2$). The effective bending stiffness($\kappa_{\text{eff}}$) can therefore be defined as the ratio between the elastic and the collisional stresses $\kappa_{\text{eff}}=\frac{\kappa}{1+g|\textbf{f}_a|^2}$ where g is a fitting parameter. If we rescale the measured persistence lengths ($\xi^\text{a}$) with the persistence length at zero activity, the data collapses onto a single curve that varies as a function of activity. Fig(~\ref{fig:fig2}b) shows that $\kappa_{\text{eff}}$ fits the collapsed curve well. We see that the activity at which the polymer persistence lengths start decreasing is where we start getting loss in segregation Fig(~\ref{fig:fig2}c). We find that this simple collision based argument provides an effective explanation as to why these active semi-flexible collectives show this reentrance.

\par

\subsection{Passive polymer properties also affect segregation dynamics}

\begin{figure}[h]
    \centering
    \includegraphics[width=\linewidth]{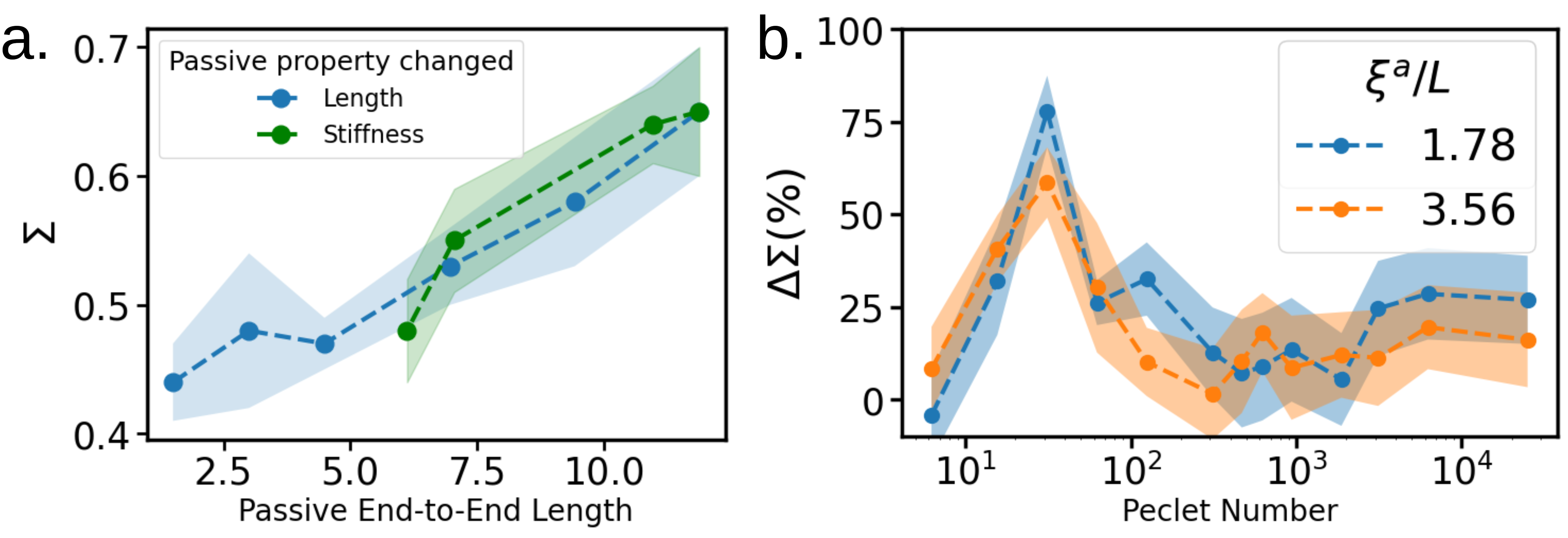}
    \caption{\textbf{Effects of the passive filament stiffness on the Segregation dynamics} (a) Plot showing the segregation order parameter as a function of the end-to-end length leads to the collapse of the data on a single line which suggests that the “effective size” of the passive filaments plays an important role in the segregation dynamics. (b) Plot showing the percent increase in the segregation order parameter on increasing $\xi^p /L$ from 0.04 ($\Sigma_{0.04}$) to 4 ($\Sigma_{4}$) as a function of activity for two active stiffness values ($\xi^a /L :\{1.78,3.56\}$). The percent increase in segregation is only significant for a certain window of activity (Pe ~ 10 - 100).}
    \label{fig:fig3}
\end{figure}

To understand the role of passive polymers in segregation, we run simulations varying the length and stiffness of the passive polymers while keeping the activity ($Pe : 125$) and active filament stiffness ($\xi^a /L : 4 $) fixed. We find that segregation in mixtures decreases monotonically with the effective size of the passive polymer, which we characterize using the end-to-end length. We can control the effective size of the polymers by changing either their length or their stiffness. Smaller passive rods can get trapped in active clusters and break them up which lowers segregation.[Fig (\ref{fig:fig3}a)]

\par

As a part of our study, we also looked at how the decrease in segregation due to the passive filament trapping varies as a function of the activity. We calculate the percent increase in the segregation order parameter $\left(\Delta\Sigma (\%) = \frac{\Sigma_{4}-\Sigma_{0.04}}{\Sigma_{4}}\times100 \right)$ on changing $\xi^p /L$ from 4 (Stiff) to 0.04 (Flexible)  as a function of the Peclet number. We find that the trapping-induced decrease in segregation is dependent on the Peclet number [Fig.(\ref{fig:fig3}b)]. More specifically we see that there exists a range of Peclet ($\sim 10-100$) in which the percent difference in segregation is significant ($\sim 70 \%$). 

\section{\label{sec:Discussions}Discussions}

\textit{Summary} - Our work provides the set of physical principles at play that drive the segregation of mixtures of active and passive semiflexible polymers. We find that increasing the active filament stiffness leads to a monotonic increase in segregation, while increasing the Peclet number shows a reentrance in segregation at high Peclet numbers, where the mixture becomes homogeneous. We show that collisions between filaments lower the persistence lengths of the active filaments which coincides with the loss in segregation. A simple scaling argument is able to capture the lowering of the persistence lengths. It is important to note that existing literature on active polymers and collections of such objects have reported similar lowering of the persistence lengths \cite{IseleHolder2015,Duman2018,Prathyusha2018}, but the lowering has been explained by the formation of spirals, which is easier for high aspect ratio filaments. We find that collision driven lowering of the persistence lengths is the dominant mode for low aspect ratio filaments. Finally, we show how the passive filament properties (length or stiffness) affect the segregation. We find that lowering the passive stiffness leads to a lowering of the segregation due to passive filaments getting trapped in active clusters. 

\par 

\textit{Testing in Experiments} - The simplicity of our model makes it possible for our findings to be tested in a variety of model experimental systems. Our results can be tested in \textit{in-vitro} experiments on actin-microtubule composites where the microtubules are active and driven by kinesin motors while the actin filaments are passive \cite{Farhadi2018,Sciortino2021,Sciortino2022}. Some current work has already shown how active microtubules can organise the passive objects \cite{Kuera2022}. Studies of herding of passive beads by single active filaments have also been performed in experiments with worms and robotic chains \cite{Sinaasappel2025} where the stiffness and the length of the active filaments control the passive herd size. In our model with multiple active filaments we find that the passive herd sizes are correlated to the active herd size \cite{Supplement2}.

\par 

\textit{Outlook} - We have shown a pathway for achieving compositional control of mixtures of active and passive filaments as a function of properties of the microscopic ingredients in the system. An interesting extension of our work would be by focusing on the effects of how the active filaments organize the passive filaments. Some recent work has shown how individual active filaments can herd passive objects into small clusters \cite{Sinaasappel2025,Prathyusha2023}, it will be instructive to see if a collection of active objects make the herding process more efficient or not. This can yield insights into understanding the organization in actin-MT composites such as those studied in \cite{Berezney2022,Quang2026}. Overall these findings can pave the way to engineering biomaterials with tunable properties.

\section{\label{sec:Acknowledgements}Acknowledgements}
C.B. was supported by the DST Inspire Fellowship while at IISC where most of this work was undertaken. A.B. was supported by DMR-2202353 and the Brandeis Center for Bioinspired Soft Materials, NSF MRSEC DMR-2011846. S.R. was supported in part by a J C Bose Fellowship and a National Science Chair of the ANRF, India and a Endowed Visiting Professorship of ICTS-TIFR.

\section{\label{sec:Code Availability} Code Availability}
Simulation and analysis scripts is available on Github. [Link to be inserted upon publication]. 

\section{\label{sec:Author Contributions}Author Contributions}
Conceptualization, C.B., A.B., S.R. ; methodology, C.B., A.B., and S.R.; investigation, C.B., A.B., and S.R.; writing – original draft, C.B., A.B., and S.R.; writing – review and editing, C.B., A.B., and S.R.; funding acquisition, A.B. and S.R.; resources, A.B. and S.R.; supervision, A.B. and S.R.

\bibliography{apssamp}

\ifarXiv
    \foreach \x in {1,...,\numbersupplementpages}
    {
        \clearpage
        \includepdf[pages={\x}]{\supplementfilename}
    }
\fi

\end{document}


\title{Supplementary Material}

\author{Chitrak Bhowmik}
\email{chitrak@brandeis.edu}
\affiliation{Martin Fischer School of Physics, Brandeis University, MS 057
Abelson-Bass-Yalem 107
Brandeis University
415 South Street
Waltham, MA 02453}
\affiliation{Department of Physics, IISER Berhampur, Transit campus (Govt. ITI Building)
Engg. School Junction, Berhampur
Odisha 760010}
\author{Aparna Baskaran}%
 \email{aparna@brandeis.edu}
\affiliation{Martin Fischer School of Physics, Brandeis University, MS 057
Abelson-Bass-Yalem 107
Brandeis University
415 South Street
Waltham, MA 02453}

\author{Sriram Ramaswamy}
 \homepage{sriram@iisc.ac.in}
\affiliation{Centre for Condensed-Matter Theory, Department of Physics, Indian Institute of Science, Bangalore 560 012, India}%

\date{\today}

\maketitle 
\tableofcontents

\section{Simulation Details}

\subsection{Parameters}

We present the ranges of parameters for the results presented in this paper. We also provide the LAMMPS simulation workflow.

\begin{enumerate}

    \item \textbf{Global paramters} 
    
    a. \textit{Simulation box size} - 200$\sigma$ \\
    b. \textit{Number of active and passive filaments} - 500 active and 500 passive \\
    c. \textit{Temperature} - The temperature sets the scale of magnitude for the random force due to collisions from solvent molecules in our system. We set the temperature to be 1 $\epsilon /k_B$ \\
    d. \textit{Damping coefficient} - We are simulating a mesoscopic system which is in general at a low Reynolds number i.e. the inertial effects are negligible. To make the system effectively overdamped we set a high damping constant ($\gamma=0.1$). Doing this is more computationally efficient than performing Brownian dynamics. \cite{Duman2018} \\
    e. \textit{Timesteps} - Timesteps are measured in units of $\sqrt{m\sigma^2/\epsilon}$. We set the timestep($\Delta t$) to be $10^{-3} \tau$. \\
    
    \item \textbf{Single Polymer parameters} 
    
    a. \textit{Mass of each bead} - 1 \\
    b. \textit{Length of a polymer} - Most of the runs reported in this work if not stated otherwise have been performed for N=25 i.e. 25 beads in a filament (both active and passive). \\
    c. \textit{Bond Stretching coefficient ($k_s$) } - We set $k_s$ to 5000$k_B T$ essentially making the filament in-extensible.  \\
    d.  \textit{Bending Stiffness} - We parse a wide range of persistence lengths for the active and passive filaments. Active persistence lengths $\xi^{\text{a}} : \{0.4,1,2,4,12\}$ , while passive persistence length $\xi^{\text{p}} : \{0.1,0.5,1,5,10,50,100\}$\\
    e. \textit{Activity} - We parse a wide range of activities for the active filaments. Activity values range from $|\mathbf{f_a}| : \{0.1,0.2,0.5,0.75,1,1.5,3,5,10,20,40\}$.
    \end{enumerate}

\section{Segregation Dynamics}

We track the simulations over time to ensure that the simulations have reached steady state. As expected, increasing the activity leads to the segregation order parameter reaching steady state quicker.

\begin{figure}[h]
    \centering
    \includegraphics[width=0.5\linewidth]{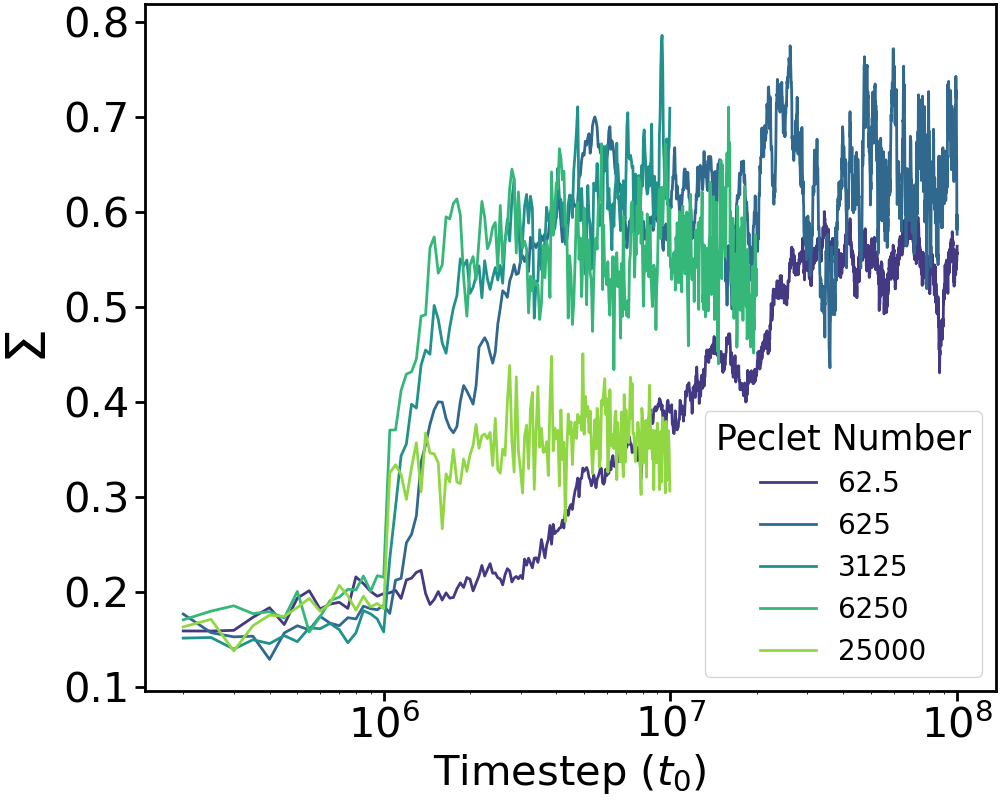}
    \caption{Timeseries of the Segregation order Parameter for different activities.}
    \label{fig:placeholder}
\end{figure}

\section{Small passive filaments destroy active clusters}

\begin{figure}[h]
    \centering
    \includegraphics[width=0.45\linewidth]{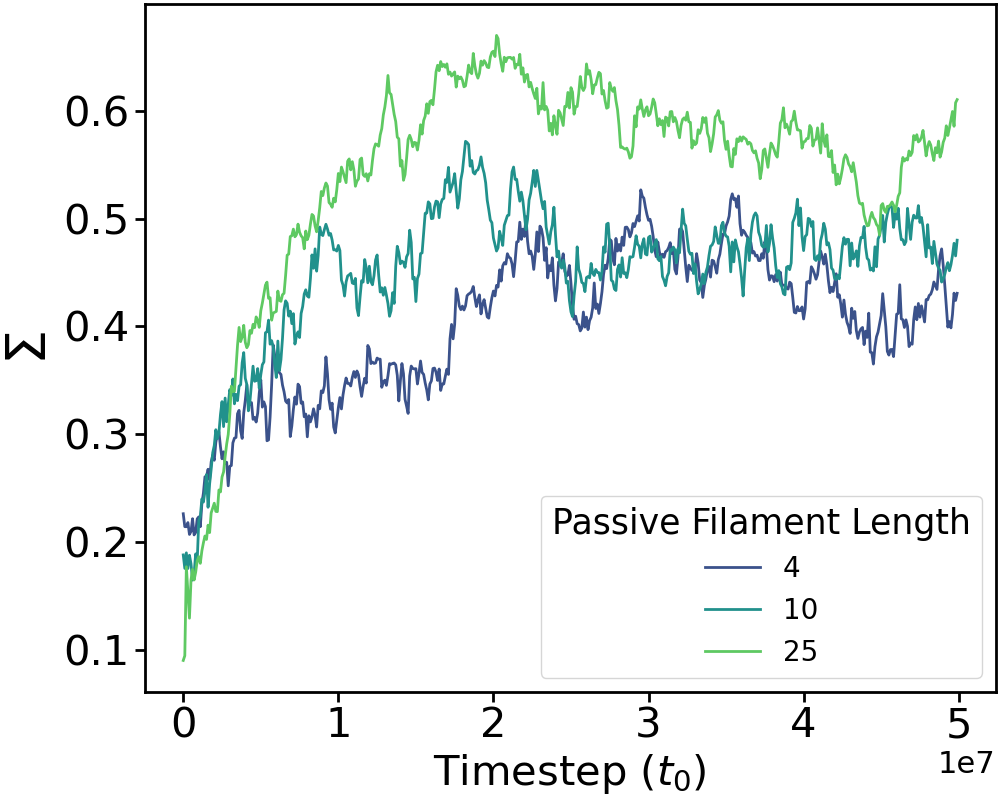}
    \includegraphics[width=0.45\linewidth]{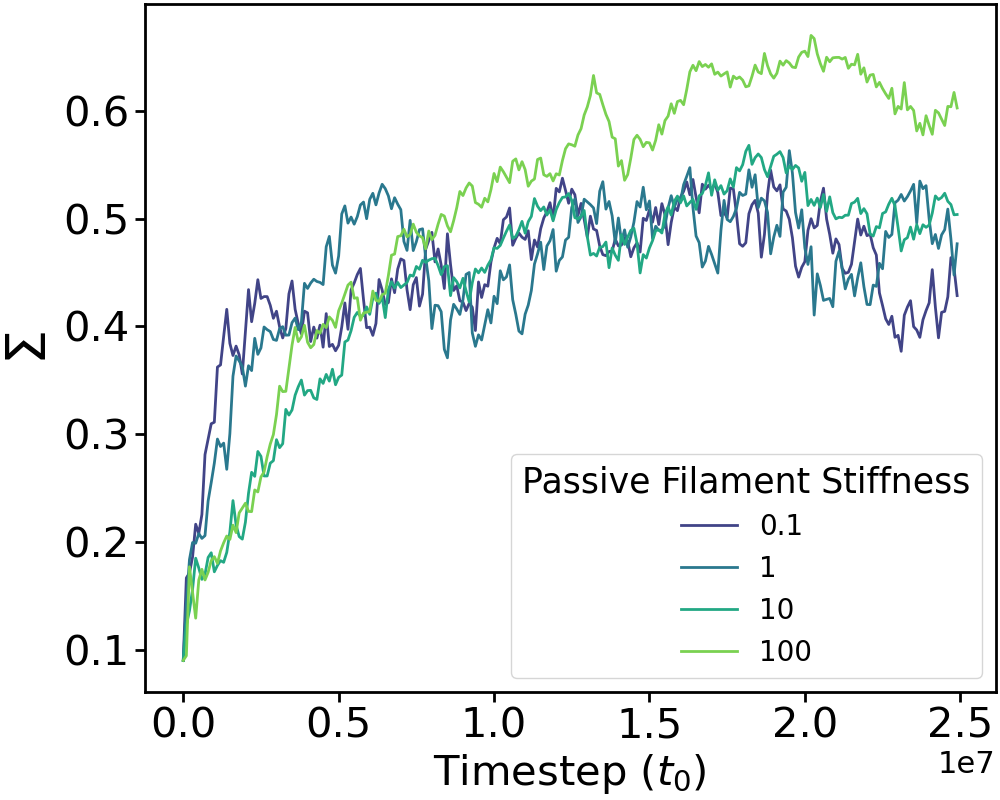}
    \caption{Timeseries of the segregation order parameter for different passive filament lengths (Left) and the passive filament stiffness (Right).}
\end{figure}

\begin{figure}
    \centering
    \includegraphics[width=0.7\linewidth]{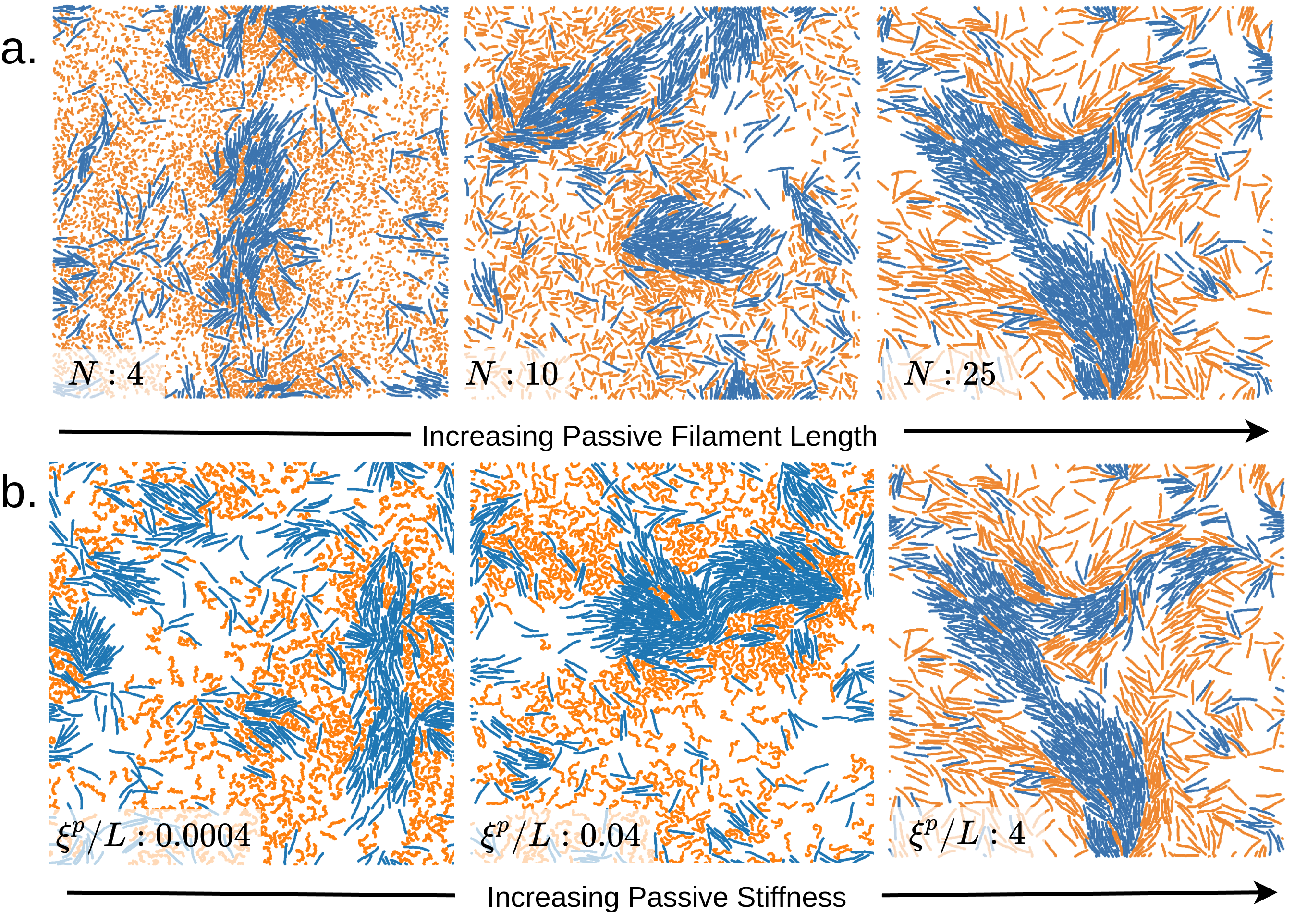}
    \caption{Snapshots showing active clusters getting broken up by smaller passive filaments, for changing passive filaments lengths and stiffness.}
    \label{fig:placeholder}
\end{figure}

\newpage

\section{Active and passive cluster sizes are correlated }
We consider two filaments to be a part of a cluster if at least $30\%$ of the filament are within a distance of $2\sigma$ and the angle between two filaments is less than $\pi/6$ \cite{Duman2018}. For each filament the algorithm involves finding the local neighbors based on the distance and angle cutoffs. Then we merge clusters that have common elements to get larger clusters. To find passive clusters we apply the same distance cutoff as in the active case. 

\begin{figure}[h]
    \centering
    \includegraphics[width=0.75\linewidth]{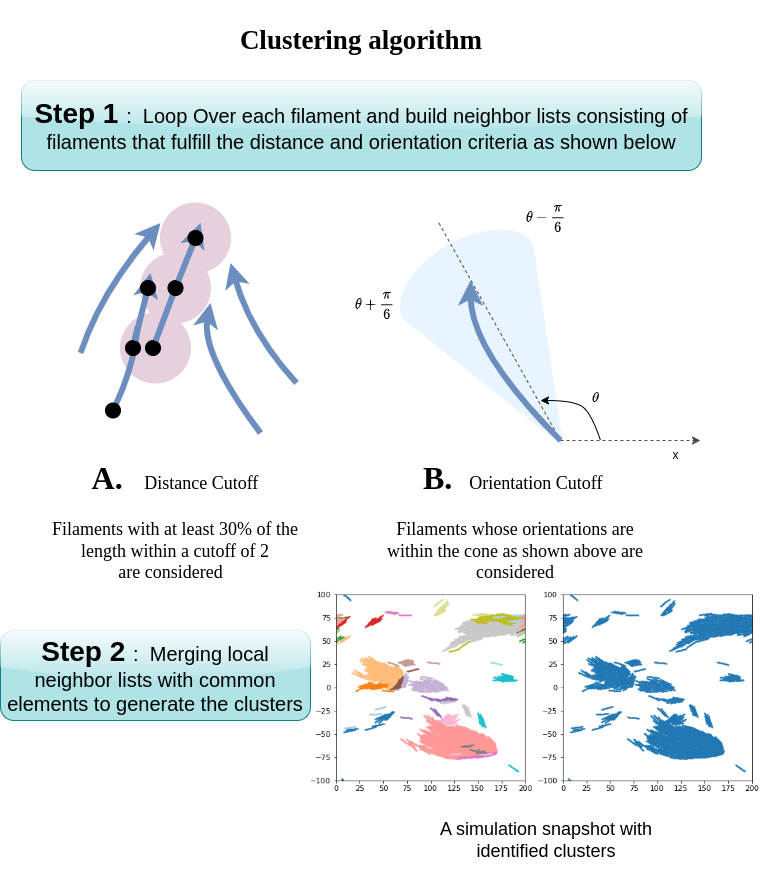}
    \caption{Schematic of the clustering algorithm for the active filaments. The distance cutoff is set to (2$\sigma$) while the orientation cutoff is set to $\pi/6$. }
    \label{fig:placeholder}
\end{figure}

We find that the active and the passive cluster sizes are correlated with each other , i.e larger active clusters organizes the passive filaments into larger herds.

\begin{figure}[h]
    \centering
    \includegraphics[width=0.75\linewidth]{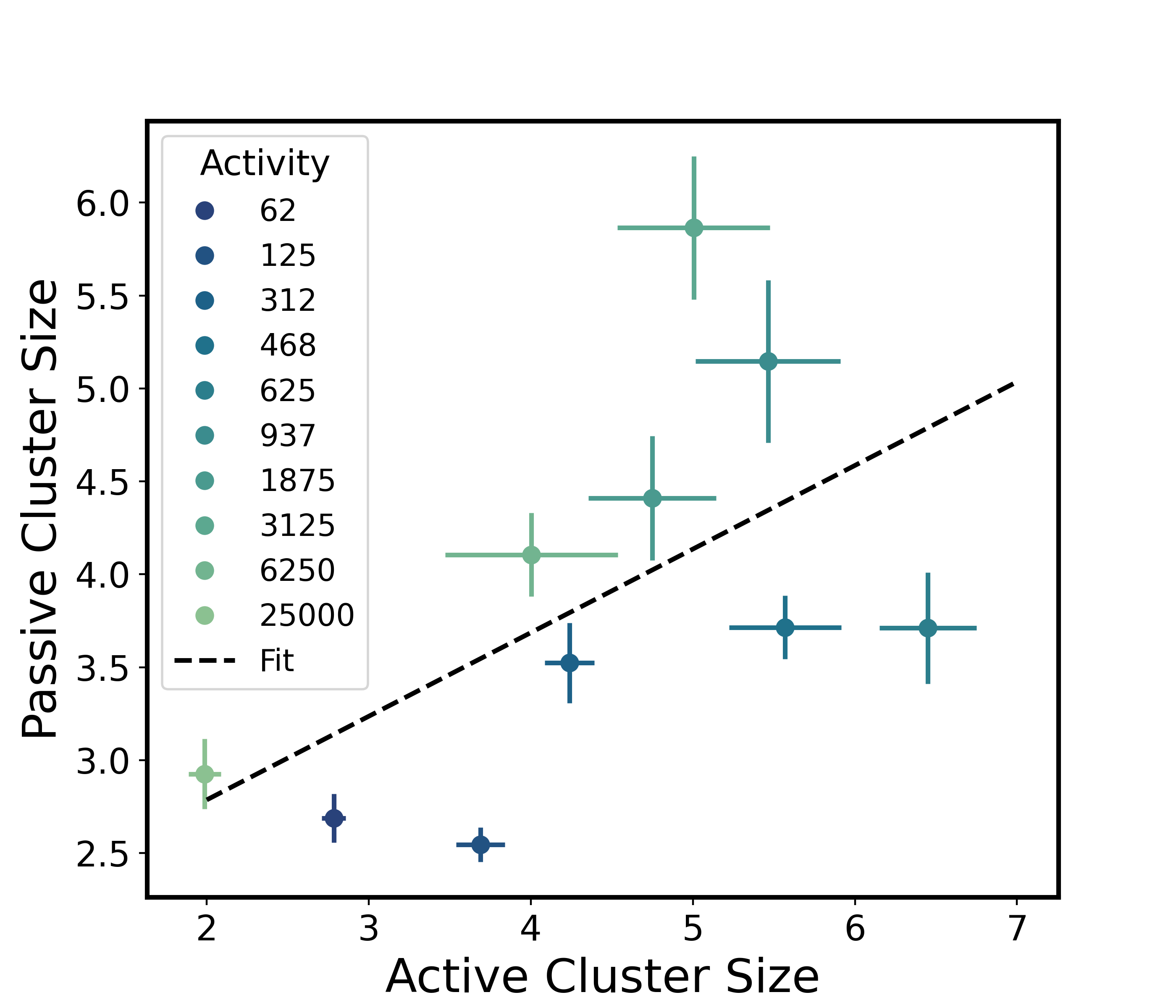}
    \caption{Plot of active cluster size vs passive cluster size for different Peclet numbers. Active  persistence length $\xi^{\text{a}}/L : 4 $ and passive persistence length $\xi^\text{p}/L : 4$.}
    \label{fig:S2}
\end{figure}

\section{Simulation Videos}

\begin{enumerate}
    \item \emph{StiffActive-IntermediatePeclet.mp4} - Peclet Number : 470 , Active $\xi^\text{a}/L : 4$, Passive $\xi^\text{p}/L : 4$  
    \item \emph{StiffActive-HighPeclet.mp4} - Peclet Number : 25000 , Active $\xi^\text{a}/L : 4$, , Passive $\xi^\text{p}/L : 4$
    \item \emph{FloppyActive-IntermediatePeclet.mp4} - Peclet Number : 470 , Active $\xi^\text{a}/L : 0.4$, , Passive $\xi^\text{p}/L : 4$
    \item  \emph{ShortPassiveFilament.mp4} - Peclet Number : 125 , Active $\xi^\text{a,p}/L : 4 $ , Passive Length : 10 beads 
    \item  \emph{FloppyPassiveFilaments.mp4} - Peclet Number : 125 , Active $\xi^\text{a}/L : 4 $ , Passive $\xi^\text{p}/L : 0.04 $
\end{enumerate}

\bibliography{apssamp}